# Comparing LLMs for Sentiment Analysis in Financial Market News

**Lucas Eduardo Pereira Teles**[1], **Carlos M. S. Figueiredo**[1]

[1]Universidade do Estado do Amazonas (UEA)
Av. Darcy Vargas – 1200 – Parque 10 de Novembro – Manaus — AM — Brasil

`{lept.eng21,cfigueiredo}@uea.edu.br`

***Abstract.*** *This article presents a comparative study of large language models (LLMs) in the task of sentiment analysis of financial market news. This work aims to analyze the performance difference of these models in this important natural language processing task within the context of finance. LLM models are compared with classical approaches, allowing for the quantification of the benefits of each tested model or approach. Results show that large language models outperform classical models in the vast majority of cases.*

## 1. Introduction

Natural Language Processing (NLP) is the process by which a computer system attempts to understand, analyze, and abstract information from human language based on a text or parts of that text [Pereira and Assunção 2022]. Sentiment analysis is one of the techniques used in the field of NLP to identify and extract information about the emotions expressed in a text, such as positivity, negativity, or neutrality [Santos et al. 2023]. This analysis can help decipher the mood and/or emotions of a general audience, as well as collect useful information about a given context, such as monitoring customer feedback on social media about the desired subject [Bollen et al. 2011].

Analyzing investor sentiment related to a particular stock can help identify patterns in stock prices [Pereira and Assunção 2022]. The goal is to understand how people feel about a particular issue or product [Santos et al. 2023]. For example, if the news mentions that "a company's profits have increased," this is clearly good news, but if the news is about "a company going bankrupt," then it is bad news [dos Santos Falcão 2020].

From a financial market perspective, such analysis can become a valuable source of information that assists investors, as it provides daily news and repercussions on their assets or investments, bringing great relevance to decision-making [de Souza et al. 2021]. LLMs (Large Language Models) are neural network models trained on a large dataset that can be used for various tasks such as sentiment analysis, text generation/summarization, etc [Bommasani et al. 2021].

To address this issue, this project consists of evaluating LLMs and other types of models for monitoring financial news, extracting information, and analyzing sentiment. Specifically, the contribution consists of: i) Collection of datasets of different types and/or formats within the financial market, such as news and reports. ii) Research on LLM techniques for efficient extraction of useful information from these data for classification. iii) Evaluation and comparison of the results of sentiment analysis of the models used.

The project is structured as follows: Section 2 shows the work that helped in the development and collection of the datasets, as well as the methods used by the authors to perform sentiment analysis. Section 3 shows the project planning, describing how the data was collected up to the pre-processing stage. Section 4 shows the results obtained from the models and analysis of each metric used for evaluation. Section 5 is reserved for final considerations and future work.

## 2. Related Works

The work [Teles and Figueiredo 2024], which studies time series consisting of the prices of six stocks, in addition to testing various models for the purpose of predicting future values. Using the iterative prediction technique, it was possible to use four types of time windows (5, 15, 25, and 35 days) to also predict the next windows. In addition to prices, texts can be interesting features to aid in prediction, which is why the current study focuses on sentiment analysis.

Analyzing sentiment in texts gives investors an idea of what the public thinks about a fund/share or a sense of whether it is worth adding it to their portfolio. It is a very powerful tool that can provide greater understanding of the subject. As discussed in [dos Santos Falcão 2020], sentiment can be extracted from an analysis of the words that compose it, with the aim of capturing a feeling of optimism or pessimism.

Choosing a model to perform this task is a very important factor. The work by [Fatemi and Hu 2023] focuses on two approaches: context-based learning and fine-tuning of LLMs, where models with fewer than 250 million to 3 billion parameters are considered for fine-tuning. Performance is compared to other state-of-the-art LLMs. The results show that smaller, fine-tuned LLMs demonstrated results comparable to current LLMs, even though they had fewer parameters and a smaller training dataset.

The dataset is another fundamental component because there is no point in having good models if the data is of poor quality. In the work by [Malo et al. 2014], the Financial Phrase Bank dataset was developed using robotic readers to analyze news texts. The goal is to investigate how semantic orientations can be better detected in financial and economic news, accommodating general information about sentence structure and the specific use of language in each domain. This dataset is used in the current work because it contains examples of financial news.

Furthermore, the work of [Lee et al. 2023] involves the creation of a new database called StockEmotions with the aim of detecting emotions in the financial market. The data for its creation was collected through a platform similar to Twitter called StockTwits. After this process, it was used for sentiment classification tasks using DistilBERT and LSTM. Only the developed dataset was used for the current work because it contains many examples of news and data already processed with emojis.

In fact, the project by [Fatemi and Hu 2023], which addressed the use of LLMs, used a dataset containing financial news headlines. This dataset was chosen for the current project in order to see how the models would perform when the data consisted of headlines rather than news articles.

The effectiveness of generative language models is a parameter that must be measured; it must be assessed whether the model performs well in the task it is performing and in the language used. The work of [de Araújo et al. 2024] investigates the learning capacity of different LLM models, evaluating their performance on a wide range of PLN datasets in Brazilian Portuguese, including three relevant sentiment analysis tasks: the identification of opinionative sentences, polarity calculation, and the identification of comparative sentences.

## 3. Methodology

### 3.1. Datasets

For the development of the project, a search was first conducted for datasets to be used. The three selected were:

- **Financial Phrase Bank (FPB):** It contains data in both Portuguese and English (which was used). Each record is a general financial news item. In total, there are 4,840 records (but the one used in the project contains 5 more records, in this case 4,845 records) containing 3 columns: the first is the "y" column, which refers to the sentiment of the news items, being neutral, positive, or negative. The second column is "text," which is the original text captured in English, and the "text_pt" column is the original text translated into Portuguese.
- **StockEmotions:** This dataset was built with the aim of detecting emotions in the financial market. It consists of 10,000 records, of which 8,000 are for training, 1,000 are for validation, and the remaining 1,000 are for testing. The sentiments in the collected dataset are bullish (positive) and bearish (negative).
- **Tweet Financial News (TFN):** The dataset contains financial news headlines collected in English. It is divided into training and validation, but only the validation data was used, which contains 2,486 records with 3 sentiments: 0 for negative, 1 for neutral, and 2 for positive.

### 3.2. Preprocessing

A major advantage of using large language models is their advanced ability to understand and extract features from natural language texts without specific preprocessing. Thus, models that follow this approach will use original news texts as test data. Classic models, on the other hand, due to the need for training, need to deal with cleaning and vectorization steps.

LLM and classic models are presented in the next section, but the preprocessing applied exclusively to classic models are:

- **Initial cleaning:** To clean up the data, links, extra spaces, and punctuation were removed, and the text was converted to lowercase.
- **Stopword removal:** After the cleaning stage, the next stage aims to remove words that appear frequently in the data but carry little or no relevant meaning. As the data is in English, stopwords belonging to the English language will be removed. To do this, the nltk package will be used to load all stopwords and apply a function to the dataset to remove them.
- **Change dataset labels:** For the FPB dataset, no change is necessary, as the labels are already in the desired format. However, StockEmotions and TFN are not. For StockEmotions, the labels are bullish and bearish, which respectively mean positive and negative. For TFN, the labels are 0, 1, and 2, which respectively mean negative, positive, and neutral. In this case, it is necessary to make the change to maintain the desired label standard.

- **Dataset balancing:** FPB and TFN are very unbalanced data sets, so it was necessary to balance them. The method used was to detect the quantity of the smallest label in the data set and collect the same quantity for the other labels, with the aim of achieving a more equal distribution of sentiments. For FPB, the number of records for each sentiment label (neutral, positive, and negative) is 2,878, 1,363, and 604, respectively. After balancing, each label had 604 records. For TFN, the number of records for each sentiment label (neutral, positive, and negative) is 1,566, 475, and 347, respectively. After balancing, each label ended up with 347 records. This process involved randomly removing records from each label using only a filter from the pandas package.
- **Division of data into training, validation, and testing:** The division of the datasets involved in the current work was to divide 20% for testing and 80% for training. Of the 80% separated, 20% was reserved for validation. For StockEmotions, it was not necessary to make the division, since the data is already separated at the time of loading, containing 8,000 records for training, 1,000 records for validation, and another 1,000 for testing. The amount of data for the FPB training, validation, and testing sets already separated are 1,159, 290, and 363, respectively. For TFN, the amounts are 665, 167, and 209, respectively.
- **Applying TF-IDF:** Using the TfidfVectorizer tool from the scikit-learn package, the TF-IDF technique consists of converting the dataset data into a TF-IDF feature matrix. TF-IDF is the product of Term Frequency (TF) and Inverse Document Frequency (IDF), which aims to find documents similar to a given search query. The vocabulary size in FPB is 5342, in StockEmotions it is 8000, and in TFN it is 2000. These values were obtained using the tool mentioned in this topic, and based on this information, matrices were generated where each record has the same size as the vocabulary.
- **Dimensionality reduction:** The final size was too large, making training unfeasible because the training time was too long. Because of this, it was necessary to reduce the size of the data using the SVD (Singular Value Decomposition) technique, and the reduction value chosen was 500 columns. It was tested with values of 250 and 1000, but with 250 the results were much lower, and with 1000 there was too much training time, which ended up exceeding the amount of memory in the environment used.

### 3.3. Models Used

### 3.3.1. Classic Models:

- **RandomForest:** Used from the scikit-learn package, it is a machine learning model belonging to the ensemble family, and the aim is to train multiple trees to solve the same problem. Their responses will be combined to obtain a better result than any individual model [Pinto and de Oliveira Nunes 2024]. Three hyperparameters were used: n_estimators, which is the number of trees used, with values of 100, 200, 300, 400, and 500; criterion, which is used to measure the quality of the tree division, with past values of "gini," "entropy," and "log loss";

and max_feature, which is the feature to be considered when searching for the best division, with past values of "sqrt," "log2," and "None." Table 1 shows the combination of values that generated the best division. Max_feature is the feature to be considered when searching for the best division, whose values were "sqrt", "log2", and "None". Table 1 shows the combination of values that generated the best results for each dataset.

Table 1. Best hyperparameters defined for Random Forest

| Dataset | n_estimators | criterion | max_features |
|---|---|---|---|
| FPB | 300 | entropy | None |
| StockEmotions | 400 | entropy | sqrt |
| TFN | 400 | gini | sqrt |

- **SVM:** The model used is the Support Vector Classifier offered by the scikit-learn package. It is a representation of examples as points in space, mapped so that the examples of each class are divided by a clear space that is as wide as possible [de Oliveira Melo and Cortes 2021]. Four hyperparameters were used: "C," which is the regularization parameter; the degree, which is the non-negative degree of the polynomial function; gamma, which is the kernel coefficient; and the kernel, which is the type to be specified in the algorithm. Table 2 shows the combination of values that generated the best results for each dataset.

Table 2. Best hyperparameters defined for SVM

| Dataset | C | Degree | Gamma | Kernel |
|---|---|---|---|---|
| FPB | 1.2 | 2 | scale | rbf |
| StockEmotions | 1.1 | 2 | scale | rbf |
| TFN | 0.7 | 2 | scale | sigmoid |

- **MLP:** Multi-Layer Perceptron is a neural network model that contains hidden layers, which can be one or more layers. The hyperparameters used were Dropout, which randomly deletes connections between layers. Epochs determines the number of times the training runs. Activation is the activation function of the layer. Batch Size is the size of the batch used in training, which was set to 32. In the last layer, there are three neurons, and using the softmax activation function, it is possible to determine the probability of a given text having a certain sentiment. For model training, methods were used to monitor training, where the maximum epoch defined for training was 200, but the training did not reach this epoch value. The monitoring techniques used were all from the Keras framework, with EarlyStopping stopping training when there is no improvement after a given number of epochs, where the value used was 10 epochs. ModelCheckpoint saves the best model during training, and ReduceLROnPlateau reduces the learning rate when the model stops improving, where the reduction value should fall by half every 5 epochs. Even when setting a default value of 200 neurons, no training reached this amount, and the number of epochs ended up being less than 20. In Table 3, the information (200, 100) means that the model has two layers of neurons containing 200 in the first and 100 in the second, with the activation function being relu in the first and selu in the second, and dropout of 0.1 in the first and none in the second. For (250), it means that the model has one layer of neurons containing 250 neurons, all of which are hidden layers.

Table 3. Best hyperparameters defined for MLPs

| Dataset | Neurônios | Ativação | Épocas | Dropout |
|---|---|---|---|---|
| FPB | (200, 100) | (relu, selu) | 13 | (0.1, 0) |
| StockEmotions | (250) | relu | 15 | 0.1 |
| TFN | (50) | elu | 16 | 0.3 |

### 3.3.2. LLM models:

- **Gemma:** A family of lightweight, open-source, state-of-the-art models developed by Google. These are broad text-to-text language models and decoders only. Gemma models are suitable for a variety of text generation tasks, including responses, summarization, reasoning, and classification, which the current project addresses. The type used is gemma-2-2b-it, where the prompt given was "Classify the text as positive, neutral, or negative. The sentiment of the text is:" The response is processed to return only the desired sentiment. For the StockEmotions dataset, the prompt given has a slight difference, as the labels are only positive and negative, so neutral is removed from the prompt.
- **DeBERTa:** Tokenizer model [He et al. 2021] improved for classification. It was trained on more than 1 million reviews from the Amazon dataset and 4 other datasets. The model has 3 labels: negative, positive, and neutral. For StockEmotions, only negative and positive were chosen.
- **DeBERTaV3:** It is an improvement on the original DeBERTa, replacing mask language modeling with token detection. The model was pre-trained using the same settings as the original to demonstrate its performance across a wide range of natural language tasks [He et al. 2023]. Different labels can be passed to the model for classification. As the models described before, we used the desired project labels, the list ['negative', 'neutral', 'positive'].
- **XLM-RoBERTa:** It is an improved version of xlm-roberta-large and, by adjusting it based on combinations of NLI data in 15, it was created to be used in zero-shot text classification [Conneau et al. 2020].
- **BART:** Adapted for text classification, serve as Zero-shot text classification (0Shot-TC). Considering the same example given for gemma and passing the same labels, the model's response is ['negative', 'neutral', 'positive'.
- **Gemini:** LLM developed by Google, Gemini was used in version 2.0-flash, focuses on content moderation and gender disparities, whereas in the current project a prompt is used to classify sentiment. The prompt used was "Classify the text as positive, neutral, or negative:". For StockEmotions, the prompt was again slightly different, with only the neutral sentiment removed.

## 4. Results

This section aims to show the results obtained from the sentiment analysis performed on the models. For each one, the three datasets were used for training/evaluation, depending on the type of model.

For SVM and Random Forest, training was not done manually. The grid search technique is used for automatic training, testing all possible parameters chosen. On the other hand, even with the use of GPUs from the collab environment, the time was quite long, reaching almost 4 hours for the random forest. For MLP, training was done manually by adjusting hyperparameters such as the number of neurons, activation function, and number of layers, but the batch size remained constant at 32 and the number of epochs varied according to the monitoring method mentioned in the "Models Used" subsection of the Methodology section.

For the FPB dataset, Table 4 below shows the results considering the precision, recall, and f1 score metrics with Positive (Pos), Neutral (Neu), and Negative (Neg) sentiments.

**Table 4. Precision, recall, and f1-score of the models tested on the FPB dataset**

|  | Precision | | | Recall | | | F1-score | | |
| --- | --- | --- | --- | --- | --- | --- | --- | --- | --- |
| Models | Pos | Neu | Neg | Pos | Neu | Neg | Pos | Neu | Neg |
| SVM | 0.71 | 0.58 | 0.77 | 0.46 | 0.81 | 0.72 | 0.56 | 0.68 | 0.74 |
| Random Forest | 0.63 | 0.53 | 0.53 | 0.19 | 0.78 | 0.68 | 0.29 | 0.63 | 0.60 |
| MLP | 0.69 | 0.61 | 0.69 | 0.55 | 0.70 | 0.75 | 0.61 | 0.65 | 0.72 |
| Gemma | 0.44 | 0.67 | 0.95 | 0.98 | 0.02 | 0.64 | 0.61 | 0.03 | 0.77 |
| DeBERTa | 1.00 | 0.72 | 1.00 | 1.00 | 1.00 | 0.53 | 1.00 | 0.84 | 0.70 |
| DeBERTaV3 | 0.57 | 0.92 | 0.77 | 0.95 | 0.09 | 0.98 | 0.71 | 0.17 | 0.86 |
| XLM-RoBERTa | 0.48 | 0.67 | 0.82 | 0.93 | 0.06 | 0.79 | 0.63 | 0.11 | 0.81 |
| BART | 0.58 | 0.86 | 0.75 | 0.96 | 0.05 | 1.00 | 0.72 | 0.09 | 0.86 |
| Gemini | 0.90 | 0.66 | 0.96 | 0.66 | 0.90 | 0.86 | 0.77 | 0.76 | 0.91 |

Looking at the F1-Score in Table 4, we see that DeBERta and Gemini performed well overall for the three classes, maintaining a value above or equal to 70%. Looking at the Negative class, almost all models detected relatively well for this class, as balancing the labels to keep the quantity of each equal had the desired effect. However, looking more closely at DeBERTa and Gemini, the Recall for DeBERTa is poor for Negative. In this case, there were records that the model was unable to detect correctly, but this label in the dataset, before being processed, has the smallest quantity, so it is natural that it has this detection difficulty. For Gemini, the Positive class was slightly below, for the same reason mentioned above, as there were characteristics that the model was unable to identify in the data. For Precision, when DeBERTa or Gemini detects the sentiment of any class, the chances of getting it right are almost 100%, with the exception of Neutral. This class has the largest number of records, and with the undersampling approach used, good records that could better identify this label may have been left out.

Looking at the F1-Score in Table 5, the SVM, MLP, and Gemini models performed well for both classes. Compared to FPB, StockEmotions records already have a column with emojis from news items that have already been processed in the texts, and these records were used here. In addition, there is a much larger amount of data to train the classic models. Therefore, even with longer training time, there are more examples for identification, and the MLP and SVM models performed better than other models,

Table 5. Precision, recall, and f1-score of the models tested on the StockEmotions dataset

|  | Precision | | recall | | F1-Score | |
|---|---|---|---|---|---|---|
| Models | Pos | Neg | Pos | Neg | Pos | Neg |
| SVM | 0.79 | 0.75 | 0.81 | 0.73 | 0.80 | 0.74 |
| Random Forest | 0.71 | 0.72 | 0.82 | 0.58 | 0.76 | 0.64 |
| MLP | 0.78 | 0.77 | 0.83 | 0.71 | 0.80 | 0.74 |
| Gemma | 0.69 | 0.65 | 0.75 | 0.57 | 0.72 | 0.61 |
| DeBERTa | 0.61 | 0.71 | 0.90 | 0.29 | 0.73 | 0.41 |
| DeBERTaV3 | 0.66 | 0.57 | 0.66 | 0.58 | 0.66 | 0.57 |
| XLM-RoBERTa | 0.63 | 0.57 | 0.70 | 0.49 | 0.67 | 0.53 |
| BART | 0.69 | 0.57 | 0.60 | 0.65 | 0.64 | 0.61 |
| Gemini | 0.80 | 0.68 | 0.70 | 0.79 | 0.75 | 0.73 |

identifying more records than the already trained models.

Table 6. Precision, recall, and f1-score of the models tested on the TFN dataset

|  | Precision | | | Recall | | | F1-score | | |
|---|---|---|---|---|---|---|---|---|---|
| Models | Pos | Neu | Neg | Pos | Neu | Neg | Pos | Neu | Neg |
| SVM | 0.62 | 0.51 | 0.55 | 0.53 | 0.71 | 0.39 | 0.57 | 0.59 | 0.46 |
| Random Forest | 0.64 | 0.54 | 0.44 | 0.47 | 0.61 | 0.52 | 0.54 | 0.58 | 0.48 |
| MLP | 0.54 | 0.59 | 0.55 | 0.62 | 0.63 | 0.42 | 0.58 | 0.61 | 0.48 |
| Gemma | 0.52 | 1.00 | 0.80 | 0.97 | 0.01 | 0.86 | 0.68 | 0.03 | 0.83 |
| DeBERTa | 0.33 | 0.00 | 0.98 | 0.49 | 0.00 | 0.97 | 0.40 | 0.00 | 0.98 |
| DeBERTaV3 | 0.69 | 0.73 | 0.59 | 0.85 | 0.11 | 0.97 | 0.76 | 0.20 | 0.74 |
| XLM-RoBERTa | 0.50 | 0.60 | 0.69 | 0.84 | 0.04 | 0.85 | 0.62 | 0.08 | 0.76 |
| BART | 0.59 | 0.33 | 0.66 | 0.89 | 0.01 | 0.95 | 0.71 | 0.03 | 0.78 |
| Gemini | 0.76 | 0.72 | 0.88 | 0.79 | 0.66 | 0.92 | 0.78 | 0.69 | 0.90 |

For the F1-Score in Table 6, only Gemini performed well compared to the other models. As mentioned earlier, the TFN dataset is a dataset that contains financial securities, so the size of the records will be smaller and contain less information than a text/news article, making sentiment identification a challenge, as can be seen for the Neutral class for almost all models. Gemini's Recall indicates that it was able to identify the three labels well, and the chances of being correct, as shown in the precision, are above 70%. Even though they are securities, Gemini achieved good results.

Overall result of the models in terms of accuracy are presented in Table 7. The classic models achieved results close to 80%, as they had many more examples than the other datasets, giving them more data to work with and thus achieving better results. In addition, Gemini was the most consistent model. Even though it did not have the highest accuracy among the three datasets, it still achieved good results compared to the other models, maintaining an accuracy above 70%.

Table 7. Accuracy, in percentage, of the models tested on the datasets

| Models | FPB | StockEmotions | TFN |
|---|---|---|---|
| SVM | 66.115 | 77.000 | 55.024 |
| Random Forest | 54.270 | 71.200 | 53.110 |
| MLP | 65.840 | 77.500 | 55.981 |
| Gemma | 54.270 | 67.300 | 61.722 |
| DeBERTa | 86.226 | 63.100 | 47.846 |
| DeBERTaV3 | 65.840 | 62.100 | 64.115 |
| XLM-RoBERTa | 58.402 | 61.100 | 57.416 |
| BART | 65.014 | 62.600 | 61.722 |
| Gemini | 80.441 | 74.100 | 78.947 |

## 5. Final Thoughts and Future Work

The presented work compared several machine learning models in the task of sentiment analysis of financial market news texts. The importance of this work lies in providing an objective basis for model selection, as well as showing the evolution of research in the area. Based on the results and consistency, it is possible to see that LLMs such as Gemini, Gemma, and BART are good LLMs to be used in various types of data, in addition to DeBERTaV3, which is an interesting choice. Furthermore, the more data provided to the classic model, the better results they can achieve, as observed for StockEmotions.

For future work, it may be interesting to use more example records in the classic models and vary the hyperparameters even more to improve training, in addition to testing other types of preprocessing and undersampling. Furthermore, with the sentiment analysis done and knowing that Gemini is a very interesting model, it is possible to align these results with a time series forecast to better monitor stock/fund variations in order to track their movements in the financial market.

## 6. Acknowledgements

I would like to thank FAPEAM (Amazonas State Research Support Foundation) for the material provided and guidance given, and the Intelligent Systems Laboratory (LSI) for providing the workspace for the development of the project and all the support/tips from the teachers and students who contributed.